\documentclass{article}
\usepackage{latexsym}
\usepackage{amsmath,amsthm}
\usepackage{graphicx}
\newtheorem{prop}{Proposition}

\title{Nonlocal symmetries for bilinear equations and their applications}
\author{ Xing-Biao Hu$^{1}$, Sen-Yue Lou$^{2,3}$ and Xian-Min Qian$^{4}$
\\\\
$^{1}$LSEC, ICMSEC,
Academy of Mathematics and Systems Science, \\
Chinese Academy of Sciences, Beijing 100190, China\\
$^{2}$Department of Physics, Ningbo University, Zhejiang , CHINA\\
$^{3}$Department of Physics, Shanghai Jiao Tong University,
Shanghai, 200240, CHINA \\
$^{4}$Department of Physics, Shaoxing College of Arts and Sciences,
Shaoxing 312000,  CHINA}

\begin{document}

\date{}
\maketitle

\begin{abstract}
In this paper, nonlocal symmetries for the bilinear KP and bilinear
BKP equations are re-studied. Two arbitrary parameters are
introduced in these nonlocal symmetries by considering gauge
invariance of the bilinear KP and bilinear BKP equations under the
transformation $f\longrightarrow fe^{ax+by+ct}$. By expanding these
nonlocal symmetries in powers of each of two parameters, we have
derived two types of bilinear NKP hierarchies and two types of
bilinear NBKP hierarchies. An impressive observation is that
bilinear positive and negative KP and BKP hierarchies may be derived
from the same nonlocal symmetries for the KP and BKP equations.
Besides,  as two concrete examples, we have deived bilinear
B\"acklund transformations for $t_{-2}$-flow of the NKP hierarchy
and $t_{-1}$-flow of the NBKP hierarchy. All these results have made
it clear that more nice integrable properties would be found for
these obtained NKP hierarchies and NBKP hierarchies. Since KP and
BKP hierarchies have played an essential role in soliton theory, we
believe that the bilinear NKP and NBKP hierarchies will have their
right place in this field.

\end{abstract}
\section{\bf Introduction}
Symmetries and conservation laws for differential equations are the
central themes of perpetual interest in mathematical
physics\cite{O,BK}. With the development of integrable systems and
soliton theory, a variety of nonlocal symmetries have been intensely
investigated in the literature, one of which is potential
symmetries. In this paper we are concerned with another type of
nonlocal symmetries, that is so-called eigenfunction symmetries
\cite{Lou93}-\cite{LW2}. To be concrete, let us first take the KP
equation as an example to see what it means by such nonlocal
symmetries. It is known that for the KP equation
\begin{equation}
(4u_t-6uu_x-u_{xxx})_x-3\alpha^2u_{yy}=0, \qquad \alpha^2=\pm
1,\label{1}
\end{equation}
we have the following nonlocal symmetry $\sigma$ \cite{MSS} given by
\begin{equation}
\sigma =(\psi\psi^*)_x\label{2}
\end{equation}
where $\psi$ and $\psi^*$ satisfy
\begin{eqnarray}
&&\alpha \psi_y+(\partial_x^2+u-\lambda )\psi=0,\label{3}\\
&&(-4\psi_t+4\psi_{xxx}+3u_x\psi+6u\psi_x-3\alpha (\partial_x^{-1}u_y)\psi =0; \label{4}\\
&&-\alpha \psi^*_y+(\partial_x^2+u-\lambda )\psi^*=0,\label{5}\\
&& (-4\psi^*_t+4\psi^*_{xxx}+3u_x\psi^*+6u\psi^*_x+3\alpha
(\partial_x^{-1}u_y)\psi^* =0. \label{6}
\end{eqnarray}
Based on the observation that equations (\ref{3}) and (\ref{4}) (or
(\ref{5}) and (\ref{6})) constitute a Lax pair for the KP (\ref{1}),
it is natural to call nonlocal
symmetry $\sigma$ in (\ref{2}) eigenfunction symmetry.\\

Eigenfunction symmetries have played an important role in the
following topics \cite{Lou93}-\cite{OS},\cite{LW1}-\cite{LH2}:

\begin{itemize}
\item Positive and negative hierarchies
\item Symmetry constraints
\item Soliton equations with sources
\end{itemize}

For example, in \cite{Lou1}, one of authors (Lou) has derived from
eigenfunction symmetry in (\ref{2}) a hierarchy of negative KP (NKP)
equations
\begin{eqnarray}
&&u_{t_{-2n-1}}=\sum_{k=0}^n(P_kQ_{n-k})_x, \label{7}\\
&&(\partial_x^2+u+\alpha \partial_y)P_k=P_{k-1},\ k=0,\ 1,\ 2,\ ...,\ n   \label{8}\\
&&(\partial_x^2+u-\alpha \partial_y)Q_k=Q_{k-1},\ k=0,\ 1,\ 2,\
...,\ n  \label{9}
\end{eqnarray}
with $P_{-1}=Q_{-1}=0$ by expanding $\psi$ and $\psi^*$ in the
following way
\begin{eqnarray*}
&&\psi =\sum_{k=0}^n(\partial_x^2+u+\alpha
\partial_y)^{n-k}P_n\lambda^k,\\
&&\psi^* =\sum_{k=0}^n(\partial_x^2+u-\alpha
\partial_y)^{n-k}Q_n\lambda^k.
\end{eqnarray*}
In particular, if $n=0$, we have the first member of the NKP
hierarchy
\begin{eqnarray}
&&u_{t_{-1}}=(P_0Q_0)_x,  \label{10}\\
&&(\partial_x^2+u+\alpha \partial_y)P_0=0, \label{11}\\
&&(\partial_x^2+u-\alpha \partial_y)Q_0=0. \label{12}
\end{eqnarray}
Through a Miura transformation, (\ref{10})-(\ref{12}) may be
transformed into (2+1)-dimensional sinh-Gordon system\cite{Lou2}
\begin{eqnarray}
&&[\alpha \phi_{yt}+e^{-2\phi}(e^{2\phi}\phi_{xt})_x]_y=-(s_xe^{2\phi})_{xx},\label{Lou1}\\
&&[e^{2\phi}(\phi_{xt}-\frac{1}{2}Ce^{2\phi}+\frac{1}{2}Ce^{-2\phi})]_{x}+\alpha
e^{2\phi}\phi_{yt}+\alpha e^{2\phi}(e^{2\phi}s)_x=0,\label{Lou2}
\end{eqnarray} where $\alpha^2=\pm1$ and C is an arbitrary constant.
Some results have been done on (\ref{Lou1})-(\ref{Lou2})
\cite{GHW}-\cite{Lou08} . However, any further integrable properties
have not been achieved for the whole NKP hierarchy
(\ref{7})-(\ref{9}). So it would be of interest to consider some
integrable properies for (\ref{7})-(\ref{9}), one of which is
Hirota's bilinear form. A usual way to do so is to find out a
suitable dependent variable transformation for the potentials in
(\ref{7})-(\ref{9}) and then try to transform (\ref{7})-(\ref{9})
into bilinear form. For example, for the first member
(\ref{10})-(\ref{12}) of the NKP hierarchy, given the dependent
variable transformation $u=2(\ln f)_{xx}, P_0=g/f, Q_0=h/f$, we can
transform (\ref{10})-(\ref{12}) into bilinear form
\begin{eqnarray}
&&D_xD_tf\cdot f=gh,\label{R1}\\
&&(D_x^2+\alpha D_y)g\cdot f=0,\label{R2}\\
&&(D_x^2+\alpha D_y)f\cdot h=0,\label{R3}
\end{eqnarray}
where Hirota bilinear operator $D^m_yD_t^k$ is defined by \cite{HRD}
\begin{equation*}
D_x^nD_y^mD_t^k a\cdot b\equiv \left (\frac \partial {\partial
x}-\frac
\partial {\partial x'}\right )^n\left (\frac \partial {\partial
y}-\frac
\partial {\partial y'}\right )^m\left (\frac \partial {\partial
t}-\frac \partial {\partial t'}\right
)^ka(x,y,t)b(x',y',t')|_{x'=x,y'=y,t'=t}.
\end{equation*}
In particular, if we set $\alpha =i, h=g^*$ in
(\ref{R1})-(\ref{R3}), we have
\begin{eqnarray}
&&D_xD_tf\cdot f=|g|^2,\label{a1}\\
&&(D_x^2+i D_y)g\cdot f=0\label{a2}
\end{eqnarray}
which is in the Hietarinta's list of complex bilinear equations
passing Hirota's 3-soliton condition\cite{H}. In particular, if
$t=y$, the system (\ref{a1}) and (\ref{a2}) is a bilinear form for
Redekopp equations\cite{MR}.

In the following, we will not follow this line to find bilinear form
for (\ref{7})-(\ref{9}). Instead we will first consider nonlocal
symmetries for blinear equations, say bilinear KP and bilinear BKP
and then deriving bilinear negative KP and negative BKP hierarchies
directly by expanding such type of nonlocal symmetries. This idea
can be described using the following diagram. \vskip .5cm

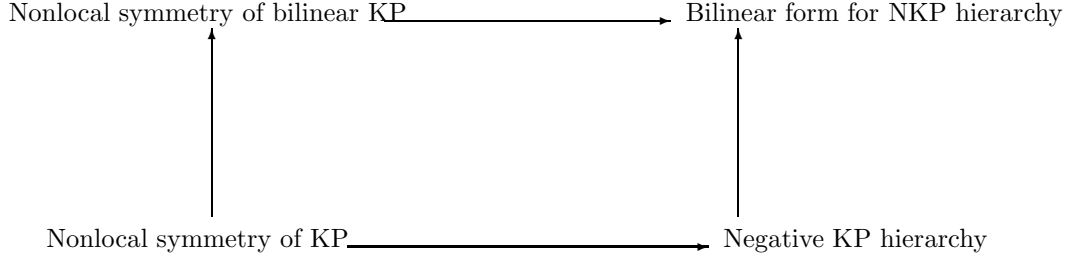
\begin{center}
\setlength{\unitlength}{1mm}
\begin{picture}(120,45)
\put(-10,10){Nonlocal symmetry of KP} \put(80,10){Negative KP
hierarchy} \put(-15,40){Nonlocal symmetry of bilinear KP}
\put(75,40){Bilinear form for NKP hierarchy}
\put(30,10){\vector(1,0){48}}\put(30,12){}
\put(35,40){\vector(1,0){38}}\put(40,42){}
\put(12,14){\vector(0,1){25}}\put(-33,24){}
\put(82,14){\vector(0,1){25}}\put(82,24){}
\end{picture}
\end{center}
According to this scheme, the problem of finding bilinear form for
the NKP hierarchy becomes the problem of constructing nonlocal
symmetry for the bilinear KP equation. One of the purposes in this
paper is to study nonlocal symmetries for bilinear equations. We
will consider nonlocal symmetries for bilinear KP equation and
bilinear BKP equation. The second purpose of the paper is to use
such nonlocal symmetries to derive bilinear forms for positive and
negative KP and BKP hierarchies. It is remarked that in \cite{LW1,
LW2} nonlocal symmetries for the bilinear KP and BKP equation have
been used to consider symmetry constraints for the KP and BKP
hierarchies.

This paper is organized as follows. In section 2, we will consider
nonlocal symmetries with two different parameters for the bilinear
KP equation and then use these symmetries to generate negative and
positive KP hierarchy in bilinear form.  Section 3 is devoted to
considering nonlocal symmetries with two different parameters for
the bilinear BKP equation and then use these symmetries to generate
negative and positive BKP hierarchy in bilinear form. Conclusions
and discussions are given in section 4.

\section{\bf Nonlocal symmetries for the bilinear KP equation and its application}
It is known that by the dependent variable transformation $u=2(\ln
f)_{xx}$, the KP equation (\ref{1}) can be transformed into the
bilinear form
\begin{equation}
(-4D_xD_t+D_x^4+3D_y^2)f\cdot f=0. \label{13}
\end{equation}
Here we have chosen $\alpha =1$ for the sake of convenience in
calculation. Concerning (\ref{13}), we have two sets of bilinear
B\"acklund transformations which are given as follows
\begin{eqnarray}
&&(D_y+D_x^2+\mu D_x-\lambda )f\cdot g=0,  \label{14}\\
&&(4D_t+3D_xD_y-D_x^3+3\mu D_y-3\lambda D_x)f\cdot g=0\label{15}
\end{eqnarray}
and
\begin{eqnarray}
&&(D_y+D_x^2+\mu D_x-\lambda )h\cdot f=0, \label{16} \\
&&(4D_t+3D_xD_y-D_x^3+3\mu D_y-3\lambda D_x)h\cdot f=0, \label{17}
\end{eqnarray}
where $\lambda$ and $\mu$ are arbitrary parameters. We have the
following result
\begin{prop}
Bilinear KP equation (\ref{13}) has a nonlocal symmetry given by
\begin{equation}
\sigma =f\int^x\frac {gh}{f^2}dx' \label{18}
\end{equation}
where $g, h$ satisfy (\ref{14})-(\ref{17}).That means $\sigma $
given by (\ref{18}) satisfies the following symmetry equation
\begin{equation}
(-4D_xD_t+3D_y^2+D_x^4)\sigma \cdot f=0.
\end{equation}
\end{prop}
\begin{proof}
By direct calculation.
\end{proof}

{\bf Remark:} This kind of nonlocal symmetry has appeared in \cite{LW1} when $\mu =0$.\\

In the following, we would like to present two sets of negative KP
hierarchies:\\

{\bf Case 1:} $$\mu =0, \  g=\sum_{i=0}^\infty g_i\lambda^i,
h=\sum_{i=0}^\infty h_i\lambda^i,$$

We have one negative KP hierarchy
\begin{eqnarray}
&&f_{t_{-2n-1}}=\frac 12\frac 1{n!}f\int^x\frac {\left (\frac
{\partial^n(gh)}{\partial\lambda^n}\right )|_{\lambda =0}}{f^2}dx'\\
&&(D_y+D_x^2-\lambda )f\cdot g=0  \\
&&(D_y+D_x^2-\lambda )h\cdot f=0
\end{eqnarray}
i.e.
\begin{eqnarray}
&&D_xD_{t_{-2n-1}}f\cdot f=\sum_{i=0}^ng_ih_{n-i}\label{19}\\
&&(D_y+D_x^2)f\cdot g_i=fg_{i-1} \label{20}\\
&&(D_y+D_x^2)h_i\cdot f=h_{i-1}f\label{21}
\end{eqnarray}
with $g_{-1}=h_{-1}=0$. Obviously, equations (\ref{19})-(\ref{21})
constitute bilinear form for the NKP hierarchy (\ref{7})-(\ref{9})
with $\alpha =1$. In consideration of the fact that there is only
one $\tau$ function appeared in the bilinear form for positive KP
hierarchy, it is natural to inquire whether we may also transform
(\ref{19})-(\ref{21}) into a set of bilinear equations with only one
$\tau$ function. The answer is affirmative. In the following,
through concrete examples, we will show that by introducing a
sequence of additional variables $m, z_1, z_2, \cdots $, equations
(\ref{19})-(\ref{21}) may be transformed into a set of bilinear
equations with one $\tau$-function.

{\bf Example 1: n=0.} In this case, we set $f=f(m), g_0=f(m-1),
h_0=f(m+1)$, Then $t_{-1}$-flow of the NKP hierarchy
(\ref{19})-(\ref{21}) becomes
\begin{eqnarray}
&&D_xD_{t_{-1}}f\cdot f=e^{D_m}f\cdot f,\\
&&(D_y+D_x^2)e^{\frac 12D_m}f\cdot f=0,
\end{eqnarray}
where Hirota's bilinear difference operator $\exp(\delta D_m)$ is
defined by
\begin{equation*}
\exp(\delta D_m)a(m)\cdot b(m)\equiv \exp\left
[\delta(\frac{\partial}{\partial m}-\frac{\partial}{\partial m'})
\right ]a(m)b(m')\mid_{m'=m}=a(m+\delta)b(m-\delta).
\end{equation*}
{\bf Example 2: n=1.} In this case, we set
$$f=f(m), g_0=f(m-1), h_0=f(m+1), g_1=-f_{z_1}(m-1), h_1=f_{z_1}(m+1).$$
Then $t_{-3}$-flow of the NKP hierarchy (\ref{19})-(\ref{21})
becomes
\begin{eqnarray}
&&D_xD_{t_{-3}}f\cdot f=D_{z_1}e^{D_m}f\cdot f\\
&&(D_y+D_x^2)e^{\frac 12D_m}f\cdot f=0,\\
&&D_{z_1}(D_y+D_x^2)e^{\frac 12D_m}f\cdot f=2e^{\frac 12D_m}f\cdot
f.
\end{eqnarray}

{\bf Example 3: n=2.}  In this case, we set
$$f=f(m), g_0=f(m-1), h_0=f(m+1), g_1=-f_{z_1}(m-1), h_1=f_{z_1}(m+1), $$
$$g_2=\frac 12f_{z_1z_1}(m-1)-f_{z_2}(m-1), h_2=\frac
12f_{z_1z_1}(m+1)+f_{z_2}(m+1).$$ Then $t_{-5}$-flow of the NKP
hierarchy (\ref{19})-(\ref{21}) becomes
\begin{eqnarray}
&&D_xD_{t_{-5}}f\cdot f=(\frac 12D_{z_1}^2+D_{z_2})e^{D_m}f\cdot f\\
&&(D_y+D_x^2)e^{\frac 12D_m}f\cdot f=0,\\
&&D_{z_1}(D_y+D_x^2)e^{\frac 12D_m}f\cdot f=2e^{\frac 12D_m}f\cdot
f,\\
&&(\frac 12D_{z_1}^2+D_{z_2})(D_y+D_x^2)e^{\frac 12D_m}f\cdot
f=D_{z_1}e^{\frac 12D_m}f\cdot f.
\end{eqnarray}

In general, along this line, we may construct bilinear equations
with one $\tau$-function for the NKP hierarchy
(\ref{19})-(\ref{21}) step by step.\\

{\bf Case 2:}$$\lambda =0, \  g=\sum_{i=0}^\infty g_i\mu^i,
h=\sum_{i=0}^\infty h_i\mu^i,$$ In this case, we have another NKP
hierarchy
\begin{eqnarray}
&&f_{t_{-n-1}}=\frac 12\frac 1{n!}f\int^x\frac {\left (\frac
{\partial^n(gh)}{\partial\mu^n}\right )|_{\mu =0}}{f^2}dx'\\
&&(D_y+D_x^2+\mu D_x )f\cdot g=0  \\
&&(D_y+D_x^2+\mu D_x )h\cdot f=0
\end{eqnarray}
i.e.
\begin{eqnarray}
&&D_xD_{t_{-n-1}}f\cdot f=\sum_{i=0}^ng_ih_{n-i}\label{22}\\
&&(D_y+D_x^2)f\cdot g_i=-D_xf\cdot g_{i-1} \label{23}\\
&&(D_y+D_x^2)h_i\cdot f=-D_xh_{i-1}\cdot f \label{24}
\end{eqnarray}
with $g_{-1}=h_{-1}=0$

Again, we may construct bilinear equations with one $\tau$-function
for the NKP hierarchy (\ref{22})-(\ref{24}) by introducing a
sequence of additional variables $m, z_1, z_2, \cdots $. Here we
just consider two simplest examples.

{\bf Example 4: n=0} In this case, we set $f=f(m), g_0=f(m-1),
h_0=f(m+1)$, Then $t_{-1}$-flow of the NKP hierarchy
(\ref{22})-(\ref{24}) becomes
\begin{eqnarray}
&&D_xD_{t_{-1}}f\cdot f=e^{D_m}f\cdot f\\
&&(D_y+D_x^2)e^{\frac 12D_m}f\cdot f=0.
\end{eqnarray}
which is the same as Example 1.\\

{\bf Example 5: n=1} In this case, we set $f=f(m), g_0=f(m-1),
h_0=f(m+1),g_1=f_{z_1}(m-1), h_1=f_{z_1}(m+1)$. Then $t_{-2}$-flow
of the NKP hierarchy (\ref{22})-(\ref{24}) become
\begin{eqnarray}
&&D_xD_{t_{-2}}f\cdot f=D_{z_1}e^{D_m}f\cdot f,\label{25}\\
&&(D_y+D_x^2)e^{\frac 12D_m}f\cdot f=0,\label{26}\\
&&[D_{z_1}(D_y+D_x^2)e^{\frac 12D_m}+2D_xe^{\frac 12D_m}]f\cdot
f=0.\label{27}
\end{eqnarray}

Furthermore, concerning equations (\ref{25})-(\ref{27}), we have the
following result:

\begin{prop}
A B\"acklund transformation  for  (\ref{25})-(\ref{27}) is
\begin{eqnarray}
&&(D_xe^{-\frac{D_m}{2}}-\lambda e^{-\frac{D_m}{2}}-\mu
e^{\frac{D_m}{2}})f\cdot g=0,\label{28}\\
&&(D_x^2+D_y-2\lambda D_x)f\cdot g=0,\label{29}\\
&& (D_xD_ze^{-\frac{D_m}{2}}+\mu D_z e^{\frac{D_m}{2}}-\lambda
D_ze^{-\frac{D_m}{2}}+\gamma
e^{\frac{D_m}{2}}+e^{-\frac{D_m}{2}})f\cdot g=0.\label{30}\\
&&(D_t-\frac 1{2\mu}D_ze^{-D_m}-\frac \gamma {4\mu^2}e^{-D_m}+\zeta
)f\cdot g=0\label{31}
\end{eqnarray}
where $\lambda ,\mu, \gamma$ and $\zeta$ are arbitrary constants,
and $z\equiv z_1, t_{-2}\equiv t$ for short.
\end{prop}
\begin{proof}
Let $f(n)$ be a solution of Eqs. (\ref{25})-(\ref{27}). What we need
to prove is that the function $g$ satisfying (\ref{28})-(\ref{31})
is another solution of Eqs. (\ref{25}) and (\ref{27}), i.e.,
\begin{eqnarray}
&&P_1\equiv [D_xD_{t_{-2}}-D_{z_1}e^{D_m}]g\cdot g=0,\label{115}\\
&&P_2\equiv (D_y+D_x^2)e^{\frac 12D_m}g\cdot g=0,\label{116}\\
&&P_3\equiv  [D_{z_1}(D_y+D_x^2)e^{\frac 12D_m}+2D_xe^{\frac
12D_m}]g\cdot g=0.
\end{eqnarray}
In analogy with the proof already given in \cite{GHW}, we know that
$P_2=0$ and $P_3=0$ hold. Thus it suffices to show that $P_1=0$. In
this regard, by using (\ref{E:A.1})-(\ref{E:A.6}),  we have
\begin{eqnarray*}
&&-P_1f^2=2D_x(D_tf\cdot g)\cdot fg-2D_z\cosh (\frac 12D_m)(e^{\frac
12D_m}f\cdot g)\cdot (e^{-\frac 12D_m}f\cdot g)\\
&&=2D_x(D_tf\cdot g)\cdot fg-\frac 2\mu D_z\cosh (\frac
12D_m)(D_xe^{-\frac
12D_m}f\cdot g)\cdot (e^{-\frac 12D_m}f\cdot g)\\
&&=2D_x(D_tf\cdot g)\cdot fg+\frac 2\mu D_xfg\cdot (D_ze^{-D_m}f\cdot g)\\
&&\qquad\quad -\frac 1\mu D_x[(D_zf\cdot g)\cdot (e^{-D_m}f\cdot
g)+fg\cdot (D_ze^{-D_m}f\cdot g)]\\
&&=2D_x[(D_t-\frac 1\mu D_ze^{-D_m})f\cdot g]\cdot fg\\
&&\qquad\quad -\frac 2\mu\sinh (\frac 12D_m)[(D_xD_ze^{-\frac
12D_m}f\cdot g)\cdot (e^{-\frac 12D_m}f\cdot g)+(D_xe^{-\frac
12D_m}f\cdot g)\cdot
(D_ze^{-\frac 12D_m}f\cdot g)]\\
&&=2D_x[(D_t-\frac 1\mu D_ze^{-D_m})f\cdot g]\cdot fg\\
&&\qquad\quad -\frac 2\mu\sinh (\frac 12D_m)\left \{[(-\mu
D_ze^{\frac 12D_m}+\lambda D_ze^{-\frac 12D_m}-\gamma e^{\frac
12D_m})f\cdot g)]\cdot (e^{-\frac 12D_m}f\cdot g)\right
.\\
&&\qquad\quad \left .+[(\lambda e^{-\frac 12D_m}+\mu e^{\frac
12D_m})f\cdot g]\cdot
(D_ze^{-\frac 12D_m}f\cdot g)\right \}\\
&&=2D_x[(D_t-\frac 1\mu D_ze^{-D_m})f\cdot g]\cdot fg\\
&&\qquad\quad +2\sinh (\frac 12D_m)[(D_ze^{\frac 12D_m}f\cdot
g)\cdot (e^{-\frac 12D_m}f\cdot g)-(e^{\frac 12D_m}f\cdot g)\cdot
(D_ze^{-\frac 12D_m}f\cdot g)]\\
&&\qquad\quad +\frac {2\gamma}\mu\sinh (\frac 12D_m)(e^{\frac
12D_m}f\cdot
g)\cdot (e^{-\frac 12D_m}f\cdot g)\\
&&=2D_x[(D_t-\frac 1\mu D_ze^{-D_m})f\cdot g]\cdot fg\\
&&\qquad\quad +2D_z\cosh (\frac 12D_m)(e^{\frac 12D_m}f\cdot g)\cdot
(e^{-\frac 12D_m}f\cdot g)+\frac {2\gamma}\mu\sinh (\frac
12D_m)(e^{\frac 12D_m}f\cdot
g)\cdot (e^{-\frac 12D_m}f\cdot g)\\
&&=2D_x[(D_t-\frac 1{2\mu} D_ze^{-D_m})f\cdot g]\cdot fg+\frac
\gamma{\mu^2}\sinh (\frac 12D_m)(D_xe^{-\frac 12D_m}f\cdot
g)\cdot (e^{-\frac 12D_m}f\cdot g)\\
&&=2D_x[(D_t-\frac 1{2\mu} D_ze^{-D_m}-\frac
\gamma{4\mu^2}e^{-D_m})f\cdot g]\cdot fg=0
\end{eqnarray*}
\end{proof}

Next, what we want to mention is that positive KP hierarchy may be
derived from the same nonlocal symmetry (\ref{18}) but with a
different expansion. Actually we may consider the following
situation:

{\bf Case 3:}$$\lambda =0, \  g=\sum_{i=0}^\infty g_i\mu^{-i},
h=\sum_{i=0}^\infty h_i\mu^{-i}.$$

In this case, we have the following KP hierarchy
\begin{eqnarray}
&&f_{t_{n-1}}=(-1)^n\frac 1{2^n}\frac 1{n!}f\int^x\frac {\left
(\frac
{\partial^n(gh)}{\partial\mu^n}\right )|_{\mu =0}}{f^2}dx'\\
&&(D_y+D_x^2+\mu D_x )f\cdot g=0  \\
&&(D_y+D_x^2+\mu D_x )h\cdot f=0
\end{eqnarray}
i.e.
\begin{eqnarray}
&&D_xD_{t_{n-1}}f\cdot f=(-1)^n\frac 1{2^{n-1}}\sum_{i=0}^ng_ih_{n-i},\label{a22}\\
&&(D_y+D_x^2)f\cdot g_{i}=-D_xf\cdot g_{i+1}, \label{a23}\\
&&(D_y+D_x^2)h_{i}\cdot f=-D_xh_{i+1}\cdot f \label{a24}
\end{eqnarray}
with $g_{0}=h_{0}=f$.

By direct calculations, we have,
$$
g_1=2f_x, g_2=2(-f_y+f_{xx}), g_3=\frac 83f_{t_3}+\frac
43f_{xxx}-4f_{xy}, \cdots
$$
and
$$
h_1=-2f_x, h_2=2(f_y+f_{xx}), h_3=-\frac 83f_{t_3}-\frac
43f_{xxx}-4f_{xy}, \cdots
$$
from which we have
$$D_xD_{t_1}f\cdot f=D_x^2f\cdot f, \qquad D_xD_{t_2}f\cdot f=D_xD_yf\cdot f$$
which means we may choose $t_1\equiv x$ and $t_2\equiv y$.

{\bf Remark} In Sato theory, there is a famous generating formula
for deriving all the bilinear equations of the KP hierarchy
\cite{DKJM}-\cite{WS}
\section{\bf Nonlocal symmetries for the bilinear BKP equation and its application}
In this section, we will consider nonlocal symmetry for bilinear BKP
equation and its application to generating negative and positive BKP
hierarchies. The bilinear BKP reads \cite{DJKM}
\begin{equation}
(D_x^6-5D_x^3D_y-5D_y^2+9D_xD_t)f\cdot f=0.\label{BKP}
\end{equation}
Its bilinear BT is given as follows:
\begin{eqnarray}
&&(D_x^3-D_y)f\cdot g=0,\label{BT1} \\
&&(D_x^5+5D_x^2D_y-6D_t)f\cdot g=0,\label{BT2}
\end{eqnarray}
or equivalently
\begin{eqnarray}
&&(D_x^3-D_y)f\cdot h=0, \label{BT3}\\
&&(D_x^5+5D_x^2D_y-6D_t)f\cdot h=0.\label{BT4}
\end{eqnarray}
We have the following result \cite{LW2}
\begin{prop}
$\sigma$ given by
$$\sigma =f\int^x\frac {D_xg\cdot h}{f^2}dx'$$ is a nonlocal
symmetry for the BKP equation (\ref{BKP}), i.e. $\sigma$ satisfies
symmetry equation
\begin{equation}
(D_x^6-5D_x^3D_y-5D_y^2+9D_xD_t)\sigma\cdot f=0.\label{BKP1}
\end{equation}
where $g$ and $h$ satisfy (\ref{BT1})-(\ref{BT4}).
\end{prop}
If $g\longrightarrow e^{\lambda y}g$, $h\longrightarrow e^{-\lambda
y}h$, we have
\begin{eqnarray}
&&(D_x^3-D_y+\lambda )f\cdot g=0 \\
&&(D_x^5+5D_x^2D_y-6D_t-5\lambda D_x^2)f\cdot g=0
\end{eqnarray}
\begin{eqnarray}
&&(D_x^3-D_y-\lambda )f\cdot h=0 \\
&&(D_x^5+5D_x^2D_y-6D_t+5\lambda D_x^2)f\cdot h=0
\end{eqnarray}
and
$$\sigma =f\int^x\frac {D_xg\cdot h}{f^2}dx'$$ is a nonlocal
symmetry. In the following, we would like to derive a hierarchy of
negative BKP equations. For this purpose, by considering
$$ g=\sum_{i=0}^\infty g_i\lambda^i,
h=\sum_{i=0}^\infty h_i\lambda^i,$$

\noindent we may write down the following  negative BKP hierarchy
\begin{eqnarray}
&&f_{t_{-3n-1}}=\frac 12\frac 1{n!}f\int^x\frac {\left (\frac
{\partial^n(D_xg\cdot h)}{\partial\lambda^n}\right )|_{\lambda =0}}{f^2}dx'\\
&&(D_x^3-D_y+\lambda )f\cdot g=0  \\
&&(D_x^3-D_y+\lambda )h\cdot f=0
\end{eqnarray}
i.e.
\begin{eqnarray}
&&D_xD_{t_{-3n-1}}f\cdot f=\sum_{i=0}^nD_xg_i\cdot h_{n-i}\label{B1}\\
&&(D_x^3-D_y )f\cdot g_i+fg_{i-1}=0 \label{B2} \\
&&(D_x^3-D_y )h_i\cdot f+h_{i-1}f=0\label{B3}
\end{eqnarray}
with $g_{-1}=h_{-1}=0$. Again, in consideration of the fact that
there is only one $\tau$ function appeared in the bilinear form for
positive BKP hierarchy, it is natural to inquire as to how to
rewrite (\ref{B1})-(\ref{B3}) into a set of bilinear equations with
only one $\tau$ function. In the following, we will give some
illustrative examples to show that by introducing a sequence of
additional variables $m, z_1, z_2, \cdots $, (\ref{B1})-(\ref{B3})
may be transformed into a set of bilinear equations with one
dependent variable $f$.

{\bf Example 6: n=0.} In this case, we set $f=f(m), g_0=f(m-1),
h_0=f(m+1)$. Then $t_{-1}$-flow of the NBKP hierarchy
(\ref{B1})-(\ref{B3}) becomes
\begin{eqnarray}
&&D_xD_{t_{-1}}f\cdot f=-D_xe^{D_m}f\cdot f\label{BB1}\\
&&(D_x^3-D_y)e^{\frac 12D_m}f\cdot f=0\label{BB2}.
\end{eqnarray}

{\bf Example 7: n=1.} In this case, we set
$$f=f(m), g_0=f(m-1), h_0=f(m+1), g_1=-f_{z_1}(m-1), h_1=f_{z_1}(m+1).$$
Then $t_{-2}$-flow of the NBKP hierarchy (\ref{B1})-(\ref{B3})
becomes
\begin{eqnarray}
&&D_xD_{t_{-4}}f\cdot f=-D_xD_{z_1}e^{D_m}f\cdot f\\
&&(D_x^3-D_y)e^{\frac 12D_m}f\cdot f=0,\\
&&D_{z_1}(D_x^3-D_y)e^{\frac 12D_m}f\cdot f+2e^{\frac 12D_m}f\cdot
f=0.
\end{eqnarray}
In general, along this line, we may rewrite the NBKP hierarchy
(\ref{B1})-(\ref{B3}) in terms of one $\tau$
function, step by step.\\
On the other hand, if $g\longrightarrow e^{kx+k^3 y+k^5t}g$,
$h\longrightarrow e^{-kx-k^3y-k^5t}h$, we have from
(\ref{BT1})-(\ref{BT4}) that
\begin{eqnarray}
&&(D_x^3-D_y-3kD_x^2+3k^2D_x )f\cdot g=0 \\
&&(D_x^5+5D_x^2D_y-6D_t-5k D_x^4+5k^2D_x^3+10k^2D_y-10kD_xD_y)f\cdot
g=0
\end{eqnarray}
\begin{eqnarray}
&&(D_x^3-D_y-3kD_x^2+3k^2D_x )h\cdot f=0 \\
&&(D_x^5+5D_x^2D_y-6D_t-5k D_x^4+5k^2D_x^3+10k^2D_y-10kD_xD_y)h\cdot
f=0,
\end{eqnarray}
and
$\sigma$ given by
\begin{equation} \sigma =f\int^x\frac {(D_x+2k)g\cdot h}{f^2}dx'\label{BB}
\end{equation}
is a nonlocal symmetry. In this case, by expanding $g$ and $h$ as
follows
$$ g=\sum_{i=0}^\infty g_ik^i,
h=\sum_{i=0}^\infty h_ik^i,$$ we may write down another negative BKP
hierarchy
\begin{eqnarray}
&&f_{t_{-n-1}}=\frac 12\frac 1{n!}f\int^x\frac {\left (\frac
{\partial^n((D_x+2k)g\cdot h)}{\partial k^n}\right )|_{k =0}}{f^2}dx'\\
&&(D_x^3-D_y-3kD_x^2+3k^2D_x )f\cdot g=0 \\
&&(D_x^3-D_y-3kD_x^2+3k^2D_x )h\cdot f=0
\end{eqnarray}
i.e.
\begin{eqnarray}
&&D_xD_{t_{-n-1}}f\cdot f=\sum_{i=0}^nD_xg_i\cdot h_{n-i}+2\sum_{i=0}^{n-1}g_{i}h_{n-1-i}\label{B4}\\
&&(D_x^3-D_y )f\cdot g_i-3D_x^2f\cdot g_{i-1}+3D_xf\cdot g_{i-2}=0 \label{B5} \\
&&(D_x^3-D_y )h_i\cdot f-3D_x^2h_{i-1}\cdot f+3D_xh_{i-2}\cdot
f=0\label{B6}
\end{eqnarray}
with $g_{-2}=g_{-1}=0, h_{-2}=h_{-1}=0$. \\
In the following, we want to show you how to rewrite
(\ref{B4})-(\ref{B6}) into bilinear equations with only one $\tau$
function through some illustrative examples:

{\bf Example 8: n=0.} In this case, we set $f=f(m), g_0=f(m-1),
h_0=f(m+1)$. Then $t_{-1}$-flow of the NBKP hierarchy
(\ref{B4})-(\ref{B6}) become
\begin{eqnarray}
&&D_xD_{t_{-1}}f\cdot f=-D_xe^{D_m}f\cdot f,\label{B7}\\
&&(D_x^3-D_y)e^{\frac 12D_m}f\cdot f=0\label{B8}
\end{eqnarray}
which coincides with (\ref{BB1}) and (\ref{BB2}).

Concerning (\ref{B7}) and (\ref{B8}), we have the following result:
\begin{prop}
A B\"acklund transformation  for  (\ref{B7}) and (\ref{B8}) is
\begin{eqnarray}
&&(D_xe^{\frac{D_m}{2}}-\lambda D_xe^{-\frac{D_m}{2}}-\mu
e^{\frac{D_m}{2}}+\lambda\mu e^{-\frac{D_m}{2}})f\cdot g=0,\label{B28}\\
&&(D_x^3-D_y-3\mu D_x^2+3\mu^2D_x+\gamma )f\cdot g=0,\label{B29}\\
&&(D_{t_{-1}}+\frac 1{2\lambda }e^{D_m}-\frac \lambda 2
e^{D_m}+\zeta )f\cdot g=0\label{B30}
\end{eqnarray}
where $\lambda ,\mu, \gamma$ and $\zeta$ are arbitrary constants.
\end{prop}
\begin{proof}
Let $f(m)$ be a solution of Eqs. (\ref{B7}) and (\ref{B8}). If we
can show that Eqs. (\ref{B28})-(\ref{B30}) guarantee that the
following two relations:
\begin{eqnarray}
&&P_1\equiv (D_xD_{t_{-1}}+D_xe^{D_m})g\cdot g=0,\label{2.5}\\
&&P_2\equiv (D_x^3-D_y)e^{\frac 12D_m}g\cdot g=0,\label{2.6}
\end{eqnarray}
hold, then Eqs. (\ref{B28})-(\ref{B30}) form a $B\ddot{a}cklund$
transformation.

In analogy with the proof already given in \cite{Hu}, we know that
$P_2=0$ holds. Thus it suffices to show that $P_1=0$. In this
regard, by using (\ref{E:A.6})-(\ref{E:A.7}),  we have
\begin{eqnarray*}
&&-P_1f^2=2D_x(D_{t_{-1}}f\cdot g)\cdot fg+2\sinh (\frac
12D_m)[(D_xe^{\frac 12D_m}f\cdot g)\cdot (e^{-\frac 12D_m}f\cdot
g)-(e^{\frac
12D_m}f\cdot g)\cdot (D_xe^{-\frac 12D_m}f\cdot g)]\\
&&=2D_x(D_{t_{-1}}f\cdot g)\cdot fg+2\sinh (\frac
12D_m)[(D_xe^{\frac 12D_m}-\lambda D_xe^{-\frac 12D_m})f\cdot
g]\cdot [(-\frac 1\lambda
e^{\frac 12D_m}+e^{-\frac 12D_m})f\cdot g]\\
&&\qquad\quad+ 2\sinh (\frac 12D_m)[\frac 1\lambda (D_xe^{\frac
12D_m}f\cdot g)\cdot (e^{\frac 12D_m}f\cdot g)+\lambda (D_xe^{-\frac
12D_m}f\cdot g)\cdot (e^{-\frac 12D_m}f\cdot g)]\\
&&=2D_x(D_{t_{-1}}f\cdot g)\cdot fg+\frac 1\lambda D_x(e^{D_m}f\cdot
g)\cdot fg-\lambda D_x(e^{-D_m}f\cdot g)\cdot fg=0.
\end{eqnarray*}
\end{proof}

{\bf Example 9: n=1.} In this case, we set
$$f=f(m), g_0=f(m-1), h_0=f(m+1), g_1=-f_{z_1}(m-1), h_1=f_{z_1}(m+1).$$
Then $t_{-2}$-flow of the NBKP hierarchy (\ref{19})-(\ref{21})
becomes
\begin{eqnarray}
&&D_xD_{t_{-2}}f\cdot f=(-D_xD_{z_1}e^{D_m}+2e^{D_m})f\cdot f\\
&&(D_x^3-D_y)e^{\frac 12D_m}f\cdot f=0,\\
&&D_{z_1}(D_x^3-D_y)e^{\frac 12D_m}f\cdot f-6D_x^2e^{\frac
12D_m}f\cdot f=0.
\end{eqnarray}

{\bf Example 10: n=2.} In this case, we set
$$f=f(m), g_0=f(m-1), h_0=f(m+1), g_1=-f_{z_1}(m-1), h_1=f_{z_1}(m+1),$$
$$g_2=-f_{z_2}(m-1)+\frac 12f_{z_1z_1}(m-1), \quad h_2=f_{z_2}(m+1)+\frac 12f_{z_1z_1}(m+1)$$
Then $t_{-3}$-flow of the NBKP hierarchy (\ref{19})-(\ref{21})
becomes
\begin{eqnarray}
&&D_xD_{t_{-3}}f\cdot f=(-D_xD_{z_2}e^{D_m}-\frac 12D_xD_{z_1}^2e^{D_m}+2D_{z_1}e^{D_m})f\cdot f\\
&&(D_x^3-D_y)e^{\frac 12D_m}f\cdot f=0,\\
&&D_{z_1}(D_x^3-D_y)e^{\frac 12D_m}f\cdot f-6D_x^2e^{\frac
12D_m}f\cdot f=0,\\
&&[D_{z_2}(D_x^3-D_y)e^{\frac 12D_m}+\frac
12D_{z_1}^2(D_x^3-D_y)e^{\frac 12D_m}-3D_x^2D_{z_1}e^{\frac
12D_m}+6D_xe^{\frac 12D_m}]f\cdot f=0.
\end{eqnarray}

In general, along this line, we may rewrite the NBKP hierarchy
(\ref{B1})-(\ref{B3}) in terms of one $\tau$
function, step by step.\\

Finally, similarly as in section 2, we want to mention that positive
BKP hierarchy may be derived from the same nonlocal symmetry
(\ref{BB}) but with a different expansion. Actually we may consider
the following situation:

{\bf Case 3:}$$ g=\sum_{i=0}^\infty g_ik^{-i}, h=\sum_{i=0}^\infty
h_ik^{-i}.$$

In this case, we have the following BKP hierarchy
\begin{eqnarray}
&&f_{t_{n-1}}=\frac 14 (-1)^n\frac 1{n!}f\int^x\frac {\left (\frac
{\partial^n((D_x+2k)g\cdot h)}{\partial k^n}\right )|_{k =0}}{f^2}dx'\\
&&(D_x^3-D_y-3kD_x^2+3k^2D_x )f\cdot g=0  \\
&&(D_x^3-D_y-3kD_x^2+3k^2D_x )h\cdot f=0
\end{eqnarray}
i.e.
\begin{eqnarray}
&&D_xD_{t_{n-1}}f\cdot f=\frac 12\sum_{i=0}^nD_xg_i\cdot h_{n-i}+\sum_{i=0}^{n+1}g_{i}h_{n+1-i}\label{a22}\\
&&(D_x^3-D_y)f\cdot g_{i}-3D_x^2f\cdot g_{i+1}+3D_xf\cdot g_{i+2}=0 \label{a23}\\
&&(D_x^3-D_y)h_{i}\cdot f-3D_x^2h_{i+1}\cdot f+3D_xh_{i+2}\cdot f=0
\label{a24}
\end{eqnarray}
with $g_{0}=h_{0}=f$.

By direct calculations, we have,
$$
g_{1}=-2f_x,h_{1}=2f_x, g_2=h_2=2f_{xx}, g_3=-\frac 23f_{y}-\frac
43f_{xxx}, h_3=\frac 23f_{y}+\frac 43f_{xxx}, g_4=h_4= \frac
43f_{xy}+\frac 23f_{xxxx}, \cdots
$$
from which we have
$$D_xD_{t_1}f\cdot f=D_x^2f\cdot f \qquad D_xD_{t_2}f\cdot f=0,\qquad D_xD_{t_3}f\cdot f=D_xD_yf\cdot f$$
which means we may choose $t_1\equiv x$ and $t_3\equiv y$, and
$f_{t_2}=0$, i.e. $f$ is $t_2$-independent.
\section{Conclusion and discussions}
In this paper, we have investigated nonlocal symmetries for the
bilinear KP and bilinear BKP equations. By expanding these nonlocal
symmetries, we have derived two types of bilinear NKP hierarchies
and two types of bilinear NBKP hierarchies. Interesting thing is
that bilinear positive and negative KP and BKP hierarchies may be
derived from the same nonlocal symmetries for the KP and BKP
equations. It still remain unclear what kind of explicit relations
will exist between the obtained two NKP hierarchies or two NBKP
hierarchies. Our study strongly suggests that these obtained NKP
hierarchies and NBKP hierarchies should have many nice integrable
properties. For example, we have given a bilinear BT for
$t_{-2}$-flow of the NKP hierarchy (\ref{22})-(\ref{24}) and a
bilinear BT for $t_{-1}$-flow of the NBKP hierarchy (\ref{B7}) and
(\ref{B8}). We can also consider bilinear BTs for other members of
these NKP and NBKP hierarchies. Furthermore, using BT
(\ref{28})-(\ref{30}) and BT (\ref{B28})-(\ref{B30}), we can obtain
soliton solutions for $t_{-2}$-flow of the NKP hierarchy
(\ref{22})-(\ref{24}) and $t_{-1}$-flow (\ref{B7})-(\ref{B8}) of the
NBKP hierarchy. As for structures of $\tau$ functions for these NKP
and NBKP hierarchies, further work needs to be done. Besides, from
Proposition 3, we know that in paticular, if we choose $h=f$, we
have $\sigma =g$ is a nonlocal symmetry for the BKP equation. Then
$\Sigma =g+Cxf$ is also a symmetry for the BKP equation (\ref{BKP}).
In this case, we may derive the following negative BKP equation from
the symmetry $\Sigma$:
\begin{eqnarray}
&&f_{t_{-1}}=g+Cxf,\\
&&(D_x^3-D_y)f\cdot g=0
\end{eqnarray}
from which we have
$$
[D_{t_{-1}}(D_y-D_x^3)+6CD_x^2]f\cdot f=0.
$$
If $C=\frac 12$, then
$$
[D_{t_{-1}}(D_y-D_x^3)+3D_x^2]f\cdot f=0
$$
which coincides with $BKP_{-1}$ given in Hirota's book \cite{HRD}.

\section*{\bf Acknowledgements}
This work was supported by the National Natural Science Foundation
of China (grant nos. 10771207, 10735030 and 90503006).
\appendix
\renewcommand{\thesection}{Appendix~\Alph{section}.}
\setcounter{equation}{0}
\renewcommand{\theequation}{\mbox{A}\arabic{equation}}
\section{Hirota bilinear operator identities.}
The following bilinear operator identities hold for arbitrary
functions $a$, $b$, $c$, and $d$.
\begin{equation}
(D_xD_ta\cdot a)b^2-a^2D_xD_tb\cdot b=2D_x(D_ta\cdot b)\cdot ab
.\label{E:A.1}
\end{equation}
\begin{equation}
(D_ze^{D_m}a\cdot a)b^2-a^2D_ze^{D_m}b\cdot b=2D_z\cosh (\frac
12D_m)(e^{\frac 12D_m}a\cdot b)\cdot (e^{-\frac 12D_m}a\cdot
b).\label{E:A.}
\end{equation}
\begin{equation}
2D_z\cosh (\frac 12D_m)(D_xa\cdot b)\cdot ab=D_x[(D_ze^{\frac
12D_m}a\cdot b)\cdot (e^{-\frac 12D_m}a\cdot b)-(e^{\frac
12D_m}a\cdot b)\cdot (D_ze^{-\frac 12D_m}a\cdot b)]
\end{equation}
\begin{eqnarray}
&&D_x[(D_za\cdot b)\cdot (e^{-D_m}a\cdot b)+ab\cdot
(D_ze^{-D_m}a\cdot
b)]\nonumber\\
&&\qquad\quad =2\sinh (\frac 12D_m)[(D_xD_ze^{-\frac 12D_m}a\cdot
b)\cdot (e^{-\frac 12D_m}a\cdot b)+(D_xe^{-\frac 12D_m}a\cdot
b)\cdot (D_ze^{-\frac 12D_m}a\cdot b)]
\end{eqnarray}
\begin{equation}
\sinh (\frac 12D_m)[(D_ze^{\frac 12D_m}a\cdot b)\cdot (e^{-\frac
12D_m}a\cdot b)-(e^{\frac 12D_m}a\cdot b)\cdot (D_ze^{-\frac
12D_m}a\cdot b)]=D_z\cosh (\frac 12D_m)(e^{\frac 12D_m}a\cdot
b)\cdot (e^{-\frac 12D_m}a\cdot b)
\end{equation}
\begin{equation}
2\sinh (\frac 12D_m)(D_xa\cdot b)\cdot ab=D_x(e^{\frac 12D_m}a\cdot
b)\cdot (e^{-\frac 12D_m}a\cdot b)  \label{E:A.6}
\end{equation}
\begin{equation}
(D_xe^{D_m}a\cdot a)b^2-a^2D_xe^{D_m}b\cdot b=2\sinh (\frac
12D_m)[(D_xe^{\frac 12D_m}a\cdot b)\cdot (e^{-\frac 12D_m}a\cdot
b)-(e^{\frac 12D_m}a\cdot b)\cdot (D_xe^{-\frac 12D_m}a\cdot b)]
\label{E:A.7}
\end{equation}
 \vskip .5cm

\end{document}